\documentclass[english,aps,prd,showpacs,twocolumn,superscriptaddress,preprintnumbers,nofootinbib]{revtex4-1}

\global\long\def\order#1{\mathcal{O}\left(#1\right)}
\global\long\def\d{\mathrm{d}}

\usepackage{amsmath}
\usepackage{amssymb}

\usepackage{slashed}
\usepackage{graphicx}
\usepackage{epsfig}

\usepackage{epstopdf}
\usepackage{esint}

\begin{document}

\title{Michel decay spectrum for a muon bound to a nucleus}
\author{Andrzej Czarnecki}
\affiliation{Department of Physics, University of Alberta, Edmonton, Alberta, Canada T6G 2G7}
\author{Matthew Dowling}
\altaffiliation[Present address: ]{Institute for Theoretical Particle Physics, KIT, D-76128 Karlsruhe, Germany.
\affiliation{Department of Physics, University of Alberta, Edmonton, Alberta, Canada T6G 2G7}
}
\author{Xavier \surname{Garcia i Tormo}}
\altaffiliation[Present address: ]{Albert Einstein Center for Fundamental Physics,
  Institut f\"ur Theoretische Physik, Universit\"at Bern,
  Sidlerstrasse 5, CH-3012 Bern, Switzerland} 
\affiliation{Department of Physics, University of Alberta, Edmonton, Alberta, Canada T6G 2G7}
\author{William J. Marciano}
\affiliation{Department of Physics, Brookhaven National Laboratory,
  Upton, New York 11973, USA}
\author{Robert Szafron}
\affiliation{Department of Physics, University of Alberta, Edmonton, Alberta, Canada T6G 2G7}

\preprint{Alberta Thy 7-14}

\begin{abstract}
  The spectrum of electrons from muons decaying in an atomic bound
  state is significantly modified by their interaction with the
  nucleus.  Somewhat unexpectedly, its first measurement, at the
  Canadian laboratory TRIUMF, differed from basic theory. We show,
  using a combination of techniques developed in atomic, nuclear, and
  high-energy physics, that radiative corrections eliminate the
  discrepancy.  In addition to solving that outstanding problem, our
  more precise predictions are potentially useful for interpreting
  future high-statistics muon experiments that aim to search for
  exotic interactions at $10^{-16}$ sensitivity.
\end{abstract}


\maketitle

Muons are very special elementary particles. They exhibit essentially
the same electroweak interactions as electrons; however, their much
larger mass ($m_\mu \simeq 207 m_e$) endows them with some important
features. Most noteworthy is the free muon decay rate which stems from
its decay mode $\mu\to e\bar \nu_e \nu_\mu$.  The differential decay
rate as a function of the electron energy
\cite{Michel:1949qe,Kinoshita:1958ru} (neglecting $m_e^2/m_\mu^2$ and
$\order{\alpha^2}$ effects \cite{Anastasiou:2005pn}) is given by
\begin{equation}
\begin{split}
  \label{eq:1}
 & \frac{\d\Gamma_{\text{free}}}{\d x}
 = \frac{G_F^2 m_\mu^5}{192\pi^3} x^2 \left( 6-4x +\frac{\alpha}{\pi} f(x) \right)
\\
 & x = \frac{2E_e }{m_\mu}\qquad 0<x\le 1,
\end{split}
\end{equation}
where $\alpha=1/137.035\,999\,173(35)$  \cite{Bouchendira:2013mpa},
$G_F$ is the Fermi constant, and $f(x)$ represents rather large,
complicated radiative corrections that can
significantly modify the electron spectrum. 
The function $f(x)$ is  explicitly given~by 


\begin{eqnarray}
&f(x) =&
 \left[ \frac{5}{3 x^2}+\frac{16 x}{3}+\frac{4}{x}+(12-8 x) \ln\left(\frac{1}{x}-1\right)-8\right]
\nonumber \\& &\times \ln\left(\frac{m_\mu}{m_e}\right)  
+(6-4 x) \left[\vphantom{\frac{1}{1}} 2 \text{Li}_2(x)-2\ln^2 (x) +  \ln (x)  \right.
\nonumber \\& &\left. + \ln (1-x) \left(3 \ln
  (x)-\frac{1}{x}-1\right)-\frac{\pi ^2}{3}-2\right]
\nonumber\\
& &+\frac{(1-x)\left[34 x^2+\left(5-34 x^2+17 x\right) \ln (x)-22
    x\right]}{3 x^2} 
\nonumber \\
& & +6(1-x) \ln (x).
\end{eqnarray}

The first term  in the equation above  is enhanced by the large logarithm
$\ln\left(\frac{m_\mu}{m_e}\right)$. These large corrections vanish when integrated
over  the  electron energy,  as expected,  due to the  Kinoshita-Lee-Nauenberg (KLN)  theorem \cite{Kinoshita:1962ur,Lee:1964is}.

The experimental lifetime of a  $\mu^+$ stopped in matter,  $\tau_\mu =
2.196\,9803(22)\times 10^{-6}$ s (the most precise lifetime
measurement for any unstable state \cite{Webber:2010zf}) determines
the strength of  weak interactions quantified by $G_F=1.1663788(7)\times10^{-5}\text{GeV}^{-2}$. Comparing it with the  fine-structure  constant and other high-precision electroweak observables,
led to predictions for the  top-quark  and  Higgs-scalar  masses, before
their discovery.  

What  happens when a $\mu^{-}$, rather than a $\mu^+$, is slowed down
in matter?  In vacuum the $\mu^+$ and $\mu^-$ lifetimes must be the
same \cite{streater2000pct}; but in matter, their decays can
appear quite different. As the $\mu^-$ loses energy and starts to come
to rest, it gets bound to nuclei of charge $Z$ due to their
attractive Coulomb potential. The $\mu^-$ quickly cascades down to the
lowest 1S atomic   orbital,  where it remains in a quantum wave function
with a momentum distribution for which its average velocity is
$\langle\beta\rangle \simeq Z\alpha$.   The decreased energy of
the bound muon causes the decay-in-orbit (DIO) rate to slow down.   In addition, the electron produced in the decay feels the same
binding interaction, which increases its wave function near the decay
region, and thus the decay probability.  Interestingly, these two
effects approximately cancel \cite{ueberall60,Czarnecki:1999yj} 
due to electromagnetic gauge invariance,  and
the difference between the overall decay rates of free and bound
$\mu^-$ is mainly due to the time dilation resulting from the bound
muon's motion. (In matter, a $\mu^-$ can also undergo capture, $\mu p
\to \nu_\mu n$, which changes its effective lifetime
\cite{Andreev:2012fj,Czarnecki:2007th}.  We do not discuss that
process here.)

 While Coulombic interactions with the nucleus do not significantly 
modify the overall DIO rate (about a 0.5\% reduction from time dilation), they do make important changes to the spectrum of decay
 electrons.

  As a result of the
muon's velocity distribution, the spectrum in Eq.~(\ref{eq:1}) is
Doppler shifted and smeared.  These effects render the radiative corrections embodied in $f(x)$ quite complicated. Although, as mentioned before, radiative corrections to the electron energy spectrum of a free muon
decay are completely known up to second order in perturbation theory
\cite{Kinoshita:1958ru,Anastasiou:2005pn}, no analysis of
radiative corrections for a bound muon decay has been performed up to
now. Methods traditionally used to calculate the DIO spectrum can not
be simply extended to allow for the inclusion of radiative
corrections. 

In addition  to the above effects, nuclear-recoil on DIO leads to a very small high-energy tail in the  spectrum,  extending all the way to electron energies $E_e \sim m_{\mu}$, well above the $E_e\sim m_{\mu}/2$ end-point energy of free muon decay at
rest. Although tiny, the DIO events near $E_e\sim m_{\mu}$ are an
important background to searches for coherent $\mu-e$ conversion
experiments that will probe for new exotic interactions at $10^{-16}$
sensitivity, four orders of magnitude beyond current bounds
\cite{Onorato:2013uka,Kuno:2013mha}. A precise understanding of DIO,
not only the very-high-energy tail, but the entire electron spectrum,
will be important for calibrating and fully exploiting the intended
sensitivity of those experiments.

Leading-order,  i.e. excluding radiative corrections,  theoretical predictions for the muon DIO spectrum,
properly incorporating the effects of the Coulomb field of the nucleus
on the muon decay, as well as the finite nuclear size, have been known
for some time \cite{Haenggi:1974hp,Watanabe:1987su,Watanabe:1993}.
TWIST, an experimental muon decay program at TRIUMF, has provided the
first precision test of those expectations for a wide range of DIO
electrons with energies 18--70~MeV~\cite{Grossheim:2009aa} using an
aluminium stopping target.  Although  general agreement was found between the TWIST measurements  \cite{Grossheim:2009aa}
and theory \cite{Haenggi:1974hp,Watanabe:1987su,Watanabe:1993}, significant deviations were observed throughout the
examined spectrum, particularly in the region around the free decay
end point ($E_e \sim m_\mu/2\sim 52$ MeV) and at low energies
18--25~MeV.  

As mentioned above, quantum electrodynamics (QED) corrections were ignored in Refs.~\cite{Haenggi:1974hp,Watanabe:1987su,Watanabe:1993}.   Remarkably, the TWIST measurements  seem to be  precise enough to be sensitive to these subtle effects. Indeed, the TWIST Collaboration
noted the need for but lack of suitable radiative corrections for
their analysis.  In this paper we provide a proper computation of radiative corrections for DIO in the aforementioned energy regions.

Why has it taken five years since the completion of the TWIST
experiment for theory to catch up? The challenge is in evaluating
radiative effects for bound particles, whose interaction with the nucleus
cannot be treated as a perturbation.

However, a similar problem has been solved in quantum chromodynamics
(QCD), in the context of  heavy-quark  decays, already 20 years ago.
Interestingly, it was noted that the necessary theoretical framework
had existed in yet another area, the formalism of deep-inelastic
lepton scattering on nuclei; in his 1995 lectures Shifman wrote ``I
see absolutely no reasons why the corresponding theory was worked out
only recently and not 20 years ago'' \cite{Shifman:1995dn}.

Our goal in this paper is to complete this cycle of theoretical
developments by applying the main ideas to what should be  a
simpler 
case, namely QED.  Toward that end, we derive a shape function that
can be convoluted with the  radiatively corrected  free decay spectrum
to approximate the effects of atomic binding. The range of validity
for that prescription should extend from roughly $m_{\mu}/2$ (the free
muon decay end point \cite{Michel:1949qe}) down to much lower energies,
regions where spectral discrepancies have been uncovered by the TWIST
Collaboration. Explaining those differences was, indeed, a major
motivation for this work.  Events with higher energy,  resulting from  nuclear-recoil  effects are very
rare, but extremely important near the DIO end point $\sim m_\mu$ where
they are a background to searches for ``new physics'' via coherent $\mu-e$
conversion in atoms \cite{Czarnecki:2011mx}. Incorporating radiative
corrections in that region is not covered by our new method and is
beyond the scope of this paper.

Following Schwinger's approach \cite{Schwinger:1989ka} to bound
states, we calculate the muon-energy shift due to the field of the
nucleus as an average value of the mass operator in the $1\mathrm{S}$
state.
The optical theorem relates its imaginary part to the muon decay rate.
Denoting the sum of momenta of  the  neutrinos by $q$ we have
\begin{eqnarray}
{\mathrm d}\Gamma = \frac{ G_F^2}{E_{1S}}
\mathrm{Im}(T_{\alpha\beta})W^{\alpha\beta}\frac{{\mathrm d}^4q}{(2\pi)^3}, 
\end{eqnarray}
where $W^{\alpha\beta}$ is the neutrino  tensor, and where  we can formally write the charged particle tensor as (we use Schwinger's notation \cite{Schwinger:1989ka} and neglect the electron mass)
\begin{eqnarray}
T^{\alpha\beta}= \left\langle 1\mathrm{S} \left| \gamma^\alpha  \frac{1}{\slashed
  \Pi-\slashed q }\gamma^\beta \right| 1\mathrm{S} \right\rangle.
\end{eqnarray}
We treat the nucleus as a static source of the electric field.
 Recoil-energy  effects can be neglected for the range of electron energies
considered here, since the recoil energy is $\delta E_{\mathrm{rec}}\sim\frac{m^2_\mu (Z\alpha)^2}{2 m_N}$, with $m_N$ denoting the nucleus mass.  (In the high-energy region of the spectrum,  recoil-energy  effects are not suppressed by  $(Z\alpha)^2$, modify the maximum allowed electron energy, and cannot be neglected~\cite{Czarnecki:2011mx}.) 

The Dirac wave function, describing the 1S state of the muon, can be
approximated in the leading $Z\alpha$ order by its large components
\cite{Szafron:2013wja}.  To separate the muon motion inside the atom
from the motion of the whole system, we rewrite the covariant
derivative as $\Pi = m_\mu v +\pi$, where $v$ is the four-velocity of
the muonic atom ($v$ is timelike and $v^2=1$) and $\pi$ describes the
residual motion of the bound muon; spatial components of $\pi$ are of  order  $m_\mu Z\alpha$, and $[\pi^\alpha,\pi^\beta]=i e F^{\alpha\beta}$. We now expand the spectrum in the region where $Q^2=(m_\mu v - q)^2\approx m_\mu^2 Z\alpha$ (in the decay of a free muon,
$Q$ would be the four-momentum of the electron; this condition
requires the produced electron to be almost on-shell). Keeping only
the leading corrections in $Z\alpha$ we get
\begin{eqnarray}
T^{\alpha\beta}= \left\langle 1\mathrm{S} \left|\gamma^\alpha  \frac{\slashed Q}{Q^2
  +2\pi \cdot Q}\gamma^\beta \right| 1\mathrm{S} \right\rangle.
\end{eqnarray}
We exploit the lightness of the electron and decompose $Q$ using a
lightlike vector  $n$,  $Q= v\cdot Q\, n + \delta Q $
with $n^2=0, \; n \cdot v =1$ \cite{Mannel:1994pm}. As long as
$E_e \gg m_\mu Z\alpha$, we can neglect the term $\pi \cdot \delta Q$,
\begin{eqnarray}
\frac{1}{\pi}\mathrm{Im}(T^{\alpha\beta})&=&
\frac{m_\mu}{2}\mathrm{Tr}\left[\gamma^\alpha  \slashed 
  Q\gamma^\beta  \right(1+\slashed v)] 
\nonumber \\ &&
\times\int {\mathrm d}\lambda\, s(\lambda) \delta(Q^2+2\lambda v\cdot Q),\label{eq:imT}
\end{eqnarray}
where $s(\lambda)$ is a QED analog of the shape function
\cite{Mannel:1994pm,Neubert:1993ch,Neubert:1993um,Bigi:1993ex} that in
our case can be explicitly evaluated using the muon's Schr\"odinger wave
function $\psi(x)$,
\begin{eqnarray}
s(\lambda) = \int {\mathrm d}^3 x \psi^\star(x) \delta(\lambda-n \cdot \pi)\psi(x).
\end{eqnarray}

Great simplification can be achieved through a judicious choice of
the electromagnetic gauge, reducing the effect of the Coulomb interaction on the electron.
In the light-cone gauge, $n\cdot A=0$, we have
\begin{eqnarray}\label{eq:sx}
s(\lambda) = \int \frac{{\mathrm d}^3 k}{(2\pi)^3} \psi_g ^\star\left(\vec{k}\right)
\delta(\lambda+\vec{n}\cdot \vec{k})\psi_g\left(\vec{k}\right), 
\end{eqnarray}
where $\psi_g\left(\vec{k}\right)$ is  the muon wave function in momentum
space calculated in the light-cone gauge.  Neglecting terms quadratic
in the smearing variable $\lambda$, 
the delta function in Eq.~(\ref{eq:imT}), describing the electron's on-shell
condition, can be rewritten as  
\begin{equation}
\delta(Q^2+2\lambda v\cdot Q) \simeq
\delta(q^2-\tilde{m}^2+2\tilde{m} \tilde{E}), 
\end{equation}
with $\tilde{E} =E_e+\lambda+\frac{(Z\alpha)^2 m_\mu}{2}$ and
$\tilde{m} = m_\mu +\lambda$.  Note that in the free muon decay, the
on-shell condition for the electron is $q^2-m_\mu^2+2m_\mu E_e=0$.
The muon mass and the electron energy can be replaced by $\tilde m$
and $\tilde E$ also in the matrix element in front of the delta
function, since this introduces a change of higher order in $Z\alpha$,
beyond our target accuracy.  

Within that accuracy, the radiative corrections can be included by
substituting a matrix element squared including virtual and real
radiation for the tree-level expression in front of the integral in
Eq.~(\ref{eq:imT}).  As  a result, the expression for the DIO spectrum becomes a convolution
of the shape function with the spectrum of the free-muon decay,
in a form familiar from heavy-quark physics \cite{DeFazio:1999sv},
\begin{equation}
\frac{{\mathrm d}\Gamma}{{\mathrm d}E_e}=\left. \int {\mathrm d}\lambda
  \, s(\lambda)\frac{{\mathrm d}\Gamma_\mathrm{free}}{{\mathrm d}z}
  \frac{{\mathrm d}z}{{\mathrm d}E_e}\right|_{z\rightarrow z(\lambda)},
\label{eq:convolution}
\end{equation}
where $\frac{{\mathrm d}\Gamma_\mathrm{free}}{{\mathrm d}z}$ denotes
the differential decay rate of a free muon, including radiative
corrections, with a daughter electron carrying energy $E_e = z
m_\mu/2$, and
\begin{eqnarray}\label{eq:z0}
z(\lambda) = \frac{2 (E_e+\lambda) + (Z\alpha)^2 m_\mu }{m_\mu+\lambda}.
\end{eqnarray}
Note that we have kept a term quadratic in $Z\alpha$, arising from the
binding energy of the muon, $E_{1\mathrm{S}}\approx
m_\mu\left(1-\frac{(Z\alpha)^2}{2}\right)$. 
This term shifts
the spectrum, since the maximum energy of the electron is $E_{1\mathrm{S}}$
rather than $m_\mu$. Around $m_\mu/2$, the derivative of the spectrum with
respect to the energy behaves like $\frac{1}{Z\alpha}$, so we need to
have the
quadratic term in order to obtain the result correct to $\order{Z\alpha}$.

Eq.~(\ref{eq:convolution}) is noteworthy in several respects.  First,
the final state characterized by the observed value of $E_e$ arises
from a superposition of contributions: the energy of the electron is
modified by the motion of the muon and by the decay electron's
interaction with the nuclear field.  The probability of observing
$E_e$ should involve a square of the sum of {\em probability
  amplitudes}; but the leading binding correction results in the sum
of probabilities.

Second, in our present QED analysis, the shape function is derived
from first principles.  This is in contrast to QCD, where it was
introduced \cite{Mannel:1994pm,Neubert:1993ch,Neubert:1993um,Bigi:1993ex}.
There, because of strong interactions, the shape function cannot be
computed. Instead it has to be modelled, and constrained from
experimental data.

Finally, the decay spectrum $\d \Gamma_{\text{free}} / \d z$ refers to a {\em free
  electron}, although we know that its interaction with the nucleus
must be accounted for. Information about this interaction is
encoded in $s(\lambda)$.  This is possible thanks to gauge invariance.
The light-cone gauge  enables us to approximately treat the electron as
a free particle.

 As previously remarked,  our analysis closely resembles, and
uses the techniques employed for, the studies of heavy-quark
decays. Separation of disparate physical scales is at the basis of the
heavy-quark expansions employed there. Therefore, it is worth noticing
the corresponding energy scales which need to be taken into account
when considering muon DIO. As is manifest from the derivations
presented above, the essential idea is, like in heavy-quark systems,
the separation of bound-state energy scales from a hard energy scale,
given by $m_{\mu}$. The typical bound-state momentum in a muonic atom
is given by $m_{\mu}Z\alpha$. Therefore, the expansion parameter in
our computation is given by $(m_{\mu}Z\alpha)/m_{\mu}=Z\alpha$, which
plays a role analogous to $\Lambda_{QCD}/m_Q$ in heavy quark effective
theory (where $\Lambda_{QCD}$ is the QCD scale, and $m_Q$ the
heavy-quark mass). However, this separation of bound-state effects
from the hard scale is no longer possible in the high-energy region of
the DIO spectrum, $E_e \sim m_\mu$. In this region, in order to
produce an on-shell electron in the final state, hard photons need to
be exchanged between the muon (or the electron) and the nucleus. Our
formalism is therefore expected to work in the energy region
$E_e < (m_{\mu}/2)+m_{\mu}Z\alpha$.  A proper treatment 
of the higher-energy part of the electron spectrum is beyond the scope
of this work.  Note also  that  the convolution formula
Eq.~(\ref{eq:convolution})   allows us to calculate only  the
dominant corrections;  i.e., it does not include all corrections of
order $(Z\alpha)^2$, and beyond. These sub-leading effects would
appear as additional, parametrically suppressed, shape
functions. Subleading shape functions have been investigated to some
extent in the context of heavy-quark decays, see
e.g. Ref.~\cite{Bosch:2004cb}, and their study is quite involved. The
difficulty  in  treating and estimating the size of these subleading effects in heavy-quark systems is mainly due to the fact that shape functions encode nonperturbative effects in the QCD case. In our QED case the shape function can be derived from first principles, and it is easier to estimate the size of the neglected subleading terms. To do that, we can (i) recompute the shape function by extracting it from a (fictitious) two-body bound decay of the muon, which, without considering radiative corrections, can be computed exactly along the lines of Ref.~\cite{GarciaiTormo:2011et}. This provides a shape function which coincides with our previous computation at order $Z\alpha$ but contains different higher-order terms. (ii) Compare the muon DIO spectrum computed exactly but without radiative corrections, i.e. along the lines of Ref.~\cite{Czarnecki:2011mx}, with the spectrum computed with our convolution formula using the Born-level free decay rate. The difference between the two spectra is due to neglected $\mathcal{O}((Z\alpha)^2)$ terms in our shape function. We have performed both checks, and found that indeed the effect of higher-order terms is always of order $(Z\alpha)^2/2$, and is never larger than $1\%$. This explicitly shows that we can safely neglect subleading shape functions to describe the available experimental data. 
The complete order $(Z\alpha)^2$ corrections to
DIO can, in principle, be incorporated by combining the more exact Coulombic treatment in Ref.~\cite{Czarnecki:2011mx} with the shape function approach to ordinary radiative corrections described in this work.
The added effect is expected to be relatively small for Al with Z=13,
but could become important for much larger Z.

\begin{figure}[htb]
\vspace*{0.3cm}
\includegraphics[width=\columnwidth]{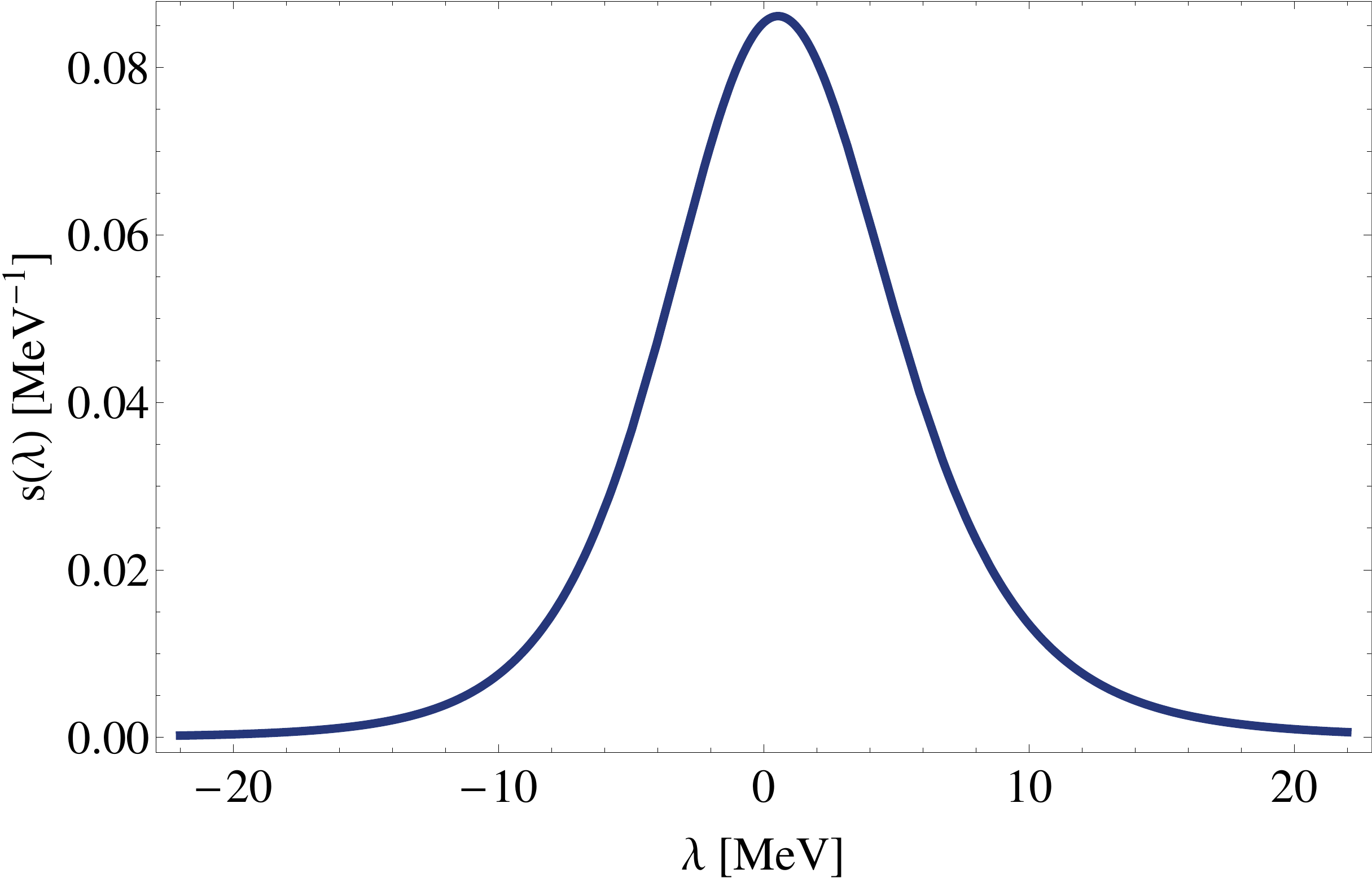}
\caption{\label{fig:sh}
 The function  $s(\lambda)$  calculated numerically for the aluminium nucleus.  The half-width of the peak is approximately $Z\alpha m_\mu
\simeq 10$ MeV.  The slight asymmetry reflects the final state interaction of
the electron.}
\end{figure}

 We also mention that finite-nuclear-size effects are more important
 in muonic atoms than in usual electronic atoms.  The muon, because of
 its relatively large mass, spends  more  of its time   close  to the nucleus.  We take into account the finite size of the nucleus when comparing  our results with 
experimental  data, and calculate the muon wave function numerically for an assumed model of charge distribution inside the
nucleus. More concretely, we  calculate the muon wave function 
as in  Refs.~\cite{Czarnecki:1998iz,Czarnecki:2011mx, GarciaiTormo:2011et},  using a  two-parameter  Fermi charge distribution: 
\begin{equation}
\varrho(r) = \frac{\varrho_0}{1+e^{\frac{r-r_0}{a}}}.
\end{equation}
In the numerical
evaluation we have focused on aluminium, $Z=13$, $r_0 =2.84 \;\text{fm}$ and $a=0.569 \; \text{fm}$
\cite{vries87}, the target
used in the TWIST experiment  \cite{Grossheim:2009aa} and 
considered as the muon stopping material for the $\mu-e$ conversion
searches at Fermilab and J-PARC \cite{Onorato:2013uka,Kuno:2013mha}.
Including binding energy, $E_{1\mathrm{S}}^{\mathrm {Al}}\approx m_\mu-0.5 \;\text{MeV}$.
Fig.~\ref{fig:sh}  shows
the function $s(\lambda)$ calculated, using
Eq.~(\ref{eq:sx}).  As expected, the main support of this function
comes from the region $\pm Z\alpha m_\mu$  around $0$, determined by the
main support of  the  muon wave function in momentum space.  We also note that uncertainties in the modeling of finite-nuclear-size effects were analyzed in Ref.~\cite{Czarnecki:2011mx}; they are not larger than our target accuracy in the present analysis, and can be safely neglected for our purposes here. 

\begin{figure}
\includegraphics[width=\columnwidth]{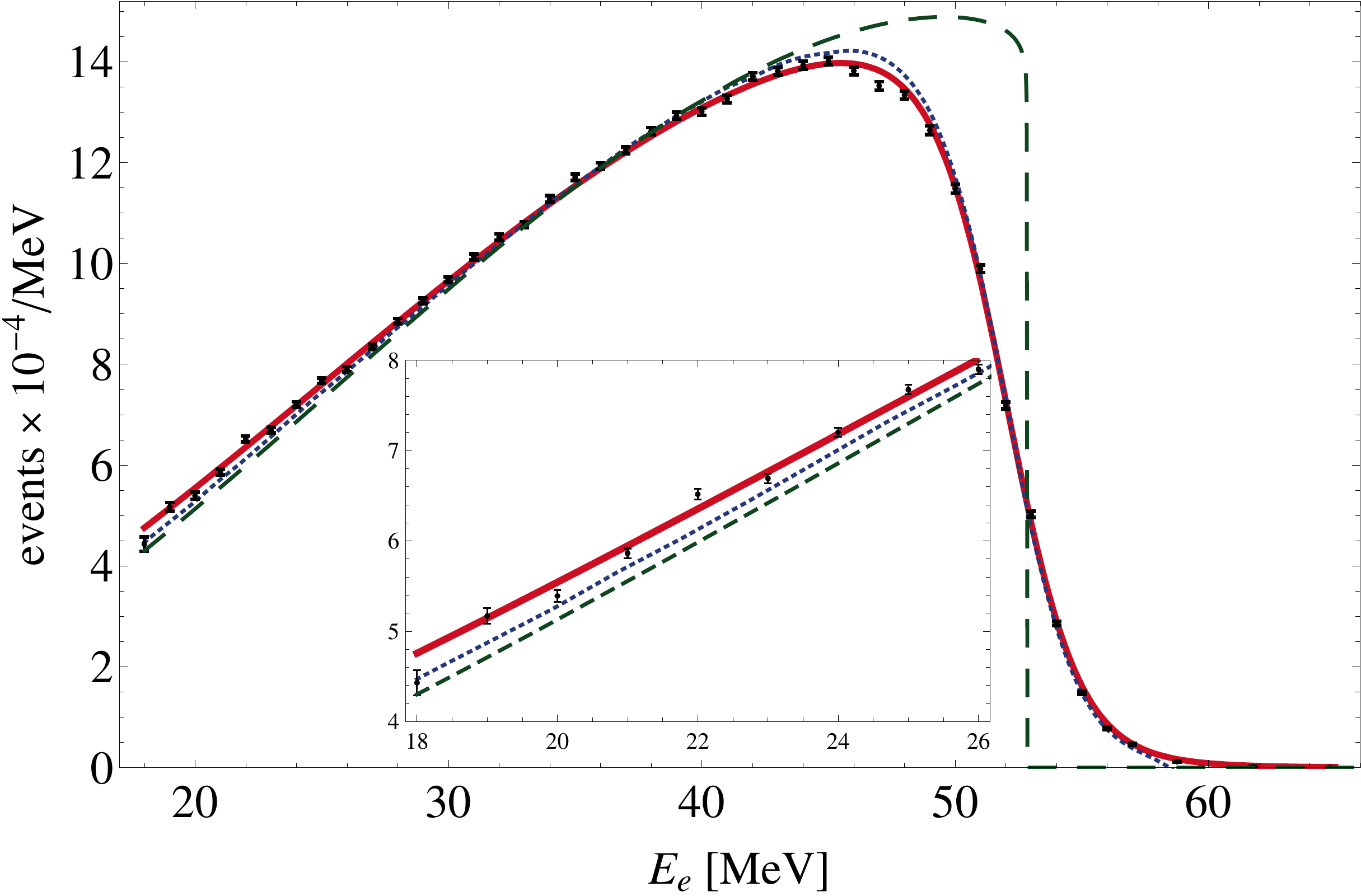}
\caption{Theoretical electron-energy spectra for
  muon DIO compared with TWIST data (black points) \cite{Grossheim:2009aa}.  The
  solid red (dotted blue) line is the spectrum with
  (without, \cite{Watanabe:1993, Czarnecki:2011mx})
  $\mathcal{O}(\alpha)$ radiative corrections.  The green dashed line
  represents the spectrum of the free muon decay with radiative
  corrections \cite{Atwood90}.  Near the free muon end point, 52.8 MeV,
  there is a large negative QED correction which pushes the dashed line
  to zero, and the solid line below the dotted one.  In the low-energy
  region, magnified in the inset, both radiative (dashed) and binding
  (dotted) corrections are positive, leading to an increase of
  low-energy electrons.  The solid line includes both
  effects. \label{fig:1} 
}
\end{figure}

The normalization of the spectrum is very important when comparing
theoretical calculations with data. The TWIST data, in addition to
statistical errors, also has an energy scale uncertainty of  $\pm 0.2\%$.   We
include it to improve the agreement between data and our
calculation by expressing the spectrum as a function $p$ of two fit
parameters $N$ and $a$,
\begin{equation}
p(N, a)=N \frac{\d\Gamma \left(a \zeta\right)}{\d\zeta} .
\label{eq:13}
\end{equation}
The parameter $a$ accounts for both experimental and theoretical
energy scale uncertainties. Its fitted value, $a \approx 1.0015$,
differs from unity within the error range claimed by TWIST,
i.e.~$\pm 2\times 10^{-3}$.

\begin{figure}
\includegraphics[width=\columnwidth]{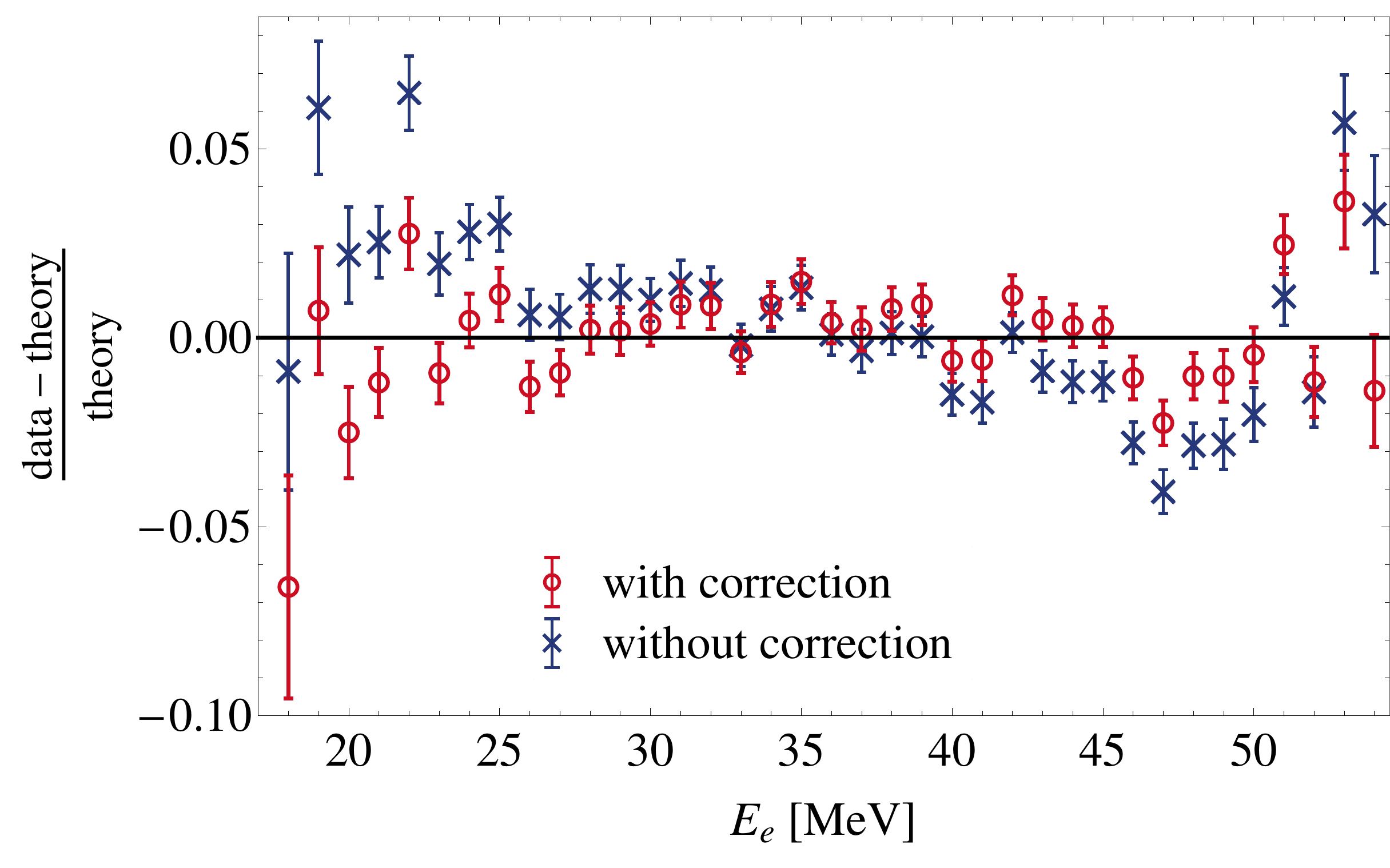}
\caption{
\label{fig:2}
Relative difference between data and the theory prediction with
(circles) and without (crosses) radiative corrections.  The difference
between the measured and the calculated spectrum is normalized to the
theoretical spectrum, appropriate for each case.  }
\end{figure}

Figure~\ref{fig:1} compares TWIST experimental data, obtained from the
decay of muons bound in aluminum with theoretical spectra (free,
lowest-order bound and including radiative corrections).  We see that
radiative corrections [obtained via Eq.~(\ref{eq:convolution})] bring theory and
experiment into good agreement. The improvement is further
demonstrated in Fig.~\ref{fig:2} which highlights the difference between theory
and experiment, with and without radiative corrections.  The rather
sizable radiative corrections (as large as 6\%) rearrange the
spectrum, but tend to cancel in the total decay rate.  We note that
improved agreement for $E_e \simeq 52-54$ MeV is due in part to a 0.15 per cent
scale shift in $a$ [see Eq.~(\ref{eq:13})] when our normalized fit to data
includes radiative corrections.  Also, the plot does not address
experimental points above 54~MeV, where our approximations  may start to fail. 

For  the  convenient use of our results, we provide a simple fit to the
spectrum which should be accurate up to effects of order
$(Z\alpha)^2\simeq 0.01$  for aluminium.  Introducing a
dimensionless variable $\zeta=\frac{2E_e}{E_{1\mathrm{S}}}$ we find
\begin{widetext}
\begin{eqnarray}
\frac{1}{\Gamma} \frac{\d\Gamma}{\d\zeta} \approx\left\{
\begin{array}{ll}
0.076+0.024 \zeta +5.92 \zeta^2 - 4.16 \zeta^3& \;\;\; 0.4<\zeta<0.76,\\
7.12 \zeta^2-5.15 \zeta+0.966 \ln (1-\zeta)+2.87& \;\;\;
0.76<\zeta<0.92, \\
\left(0.085 \zeta +1.24\right)/\left(0.714+\exp\left[36.7 (\zeta -1)\right]\right)&\;\;\; 0.92<\zeta<1.05. 
\end{array}
\right. 
\end{eqnarray}
\end{widetext}
The fit is normalized such that $\Gamma=\int_0^2 \frac{\mathrm{d}\Gamma}{\mathrm{d}\zeta}$.
The high-energy spectrum, including recoil and binding effects is
given in  Ref.~\cite{Czarnecki:2011mx};  however, radiative corrections in
that region have not been included.

\begin{figure}[htb]
\vspace*{0.3cm}
\includegraphics[width=\columnwidth]{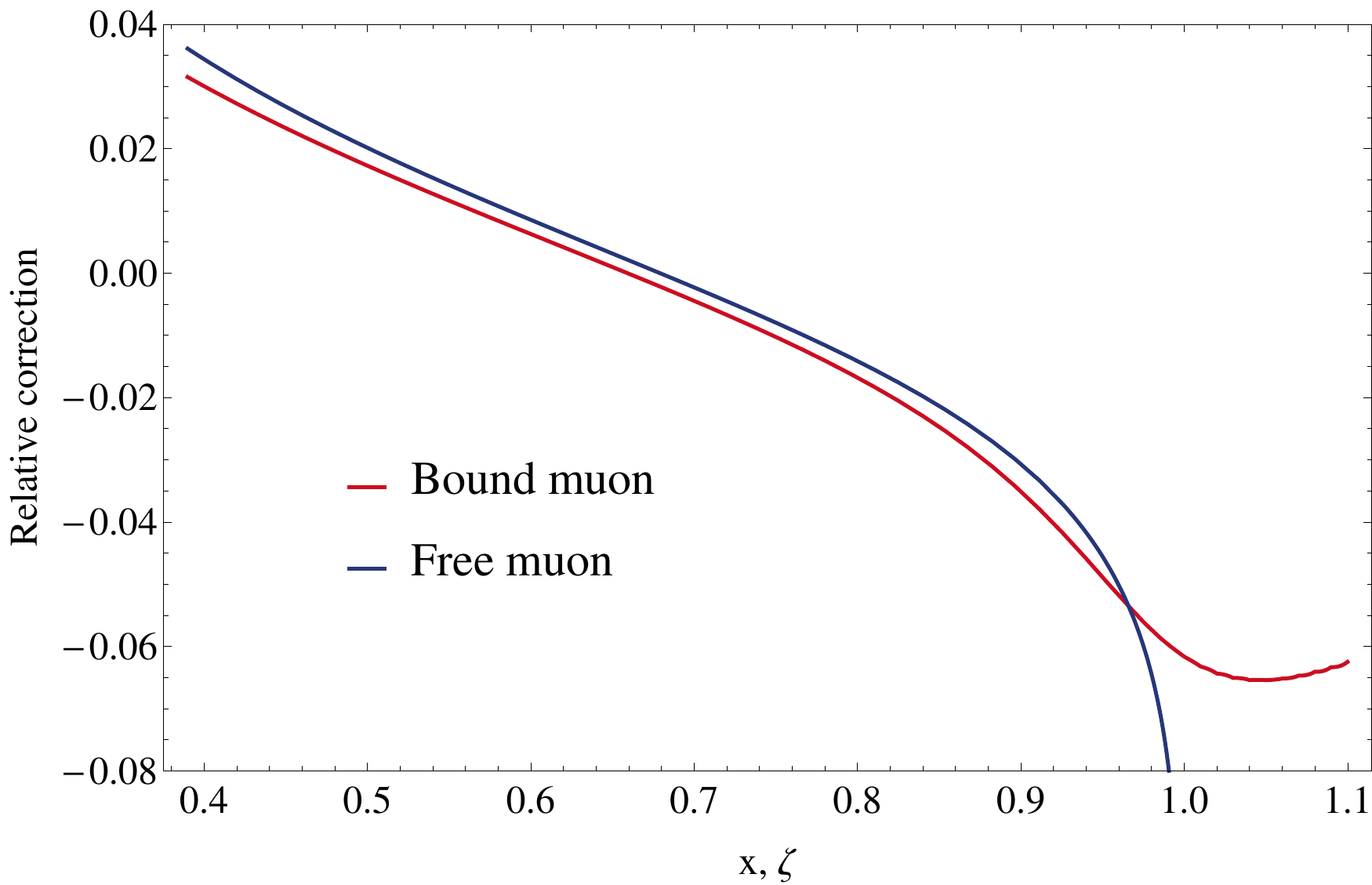}
\caption{\label{fig:relcor} Relative leading $\mathcal{O}(\alpha)$ radiative corrections to the electron spectrum for bound and free muons. The solid (dashed) line corresponds
to the corrections for a bound (free) muon as a function of the electron energy variable  $\zeta$ ($x$).}
\end{figure}

In  Fig.~\ref{fig:relcor},  the solid line describes the  $\mathcal{O}(\alpha)$ corrections to the lowest order DIO spectrum for a bound muon as a function of $\zeta$.  For comparison,  the free muon radiative corrections
are given as a function of $x$ (see  Eq.(\ref{eq:1})).  In the lower energy region, accessible to both bound and free decays, the radiative corrections are similar for the two cases, with a part of
the shift coming from a difference in the $\zeta$ and $x$
variables. Differences are largest, close to the free muon decay
spectrum end point, where a logarithmic enhancement in the free case
($\ln(1-x)$ singularity) is smeared for the DIO case, as illustrated
by the solid curve.

In summary, we have derived a new method for approximating QED
radiative corrections to muon DIO rates based on a formalism developed
for heavy-quark weak decays in QCD.  Its general features are in good
accord with expectations based on Lorenz and gauge invariance.  The
radiative corrections are quite large
near the spectral peak and at low energies, regions where the TWIST
experiment had discovered discrepancies with theory. As a result, our
new improved theoretical spectrum is now in excellent quantitative
agreement with experiment and, where applicable, can be confidently
used in future searches for exotic new physics.

\section*{Acknowledgments}
A.C., M.D., X.G.T.~and R.S.~were supported by Science and Engineering Research Canada 
(NSERC).  W.J.M.~was supported by the United States DOE under Grant No.~DE-ACO2-98CH10886.


\end{document}